\documentclass[manuscript]{aastex}

\usepackage{graphicx}
\usepackage{txfonts}
\usepackage{natbib}
\shorttitle{Plasma heating in early phase of flares}

\shortauthors{Siarkowski et al.}

\begin{document}
%


\title{Plasma heating in the very early phase of solar flares}


\author{M. Siarkowski\altaffilmark{1}}
\affil{Space Research Centre, Polish Academy of Sciences, 51-622
Wroc{\l}aw, ul. Kopernika 11, Poland}
\email{ms@cbk.pan.wroc.pl}

\author{R. Falewicz\altaffilmark{2} and P. Rudawy\altaffilmark{2}}
\affil{Astronomical Institute, University of Wroc{\l}aw, 51-622
Wroc{\l}aw, ul. Kopernika 11, Poland}
\email{falewicz@astro.uni.wroc.pl} \email{rudawy@astro.uni.wroc.pl}





\begin{abstract}
In this paper we analyze soft and hard X-ray emission of the 2002
September 20 M1.8 \emph{GOES} class solar flare observed by
\emph{RHESSI} and \textit{GOES} satellites. In this flare event,
soft X-ray emission precedes the onset of the main bulk hard X-ray
emission by $\sim$5 min. This suggests that an additional heating
mechanism may be at work at the early beginning of the flare.
However \emph{RHESSI} spectra indicate presence of the non-thermal
electrons also before impulsive phase. So, we assumed that a
dominant energy transport mechanism during rise phase of solar
flares is electron beam-driven evaporation. We used non-thermal
electron beams derived from \emph{RHESSI }spectra as the heating
source in a hydrodynamic model of the analyzed flare. We showed that
energy delivered by non-thermal electron beams is sufficient to heat
the flare loop to temperatures in which it emits soft X-ray closely
following the \emph{GOES} 1-8 \AA ~light-curve. We also analyze the
number of non-thermal electrons, the low energy cut-off, electron
spectral indices and the changes of these parameters with time.

\end{abstract}


\keywords{Sun: flares --- Sun: X-rays, gamma rays }



\section{Introduction}

It is commonly accepted that during the impulsive phase of the solar
flare, non-thermal electron beams accelerated anywhere in the solar
corona move along magnetic field lines to the chromosphere where
they deposit their energy. Here, most non-thermal electrons lose
their energy in Coulomb collisions while a small fraction ($\sim
10^{-5}$) of the electrons' energy is converted into hard X-rays
(HXR) by the bremsstrahlung process. The heated chromospheric plasma
evaporates and radiates over a wide spectral range from hard X-rays
or gamma rays to radio emission. These processes, called electron
beam-driven evaporation model is believed to be a dominant energy
transport mechanism during solar flares. One of the most spectacular
and well observed manifestation of the described processes is soft
X-ray (SXR) emission of magnetic loops observed e.g. by
\emph{Yohkoh} and \emph{Hinode} satellites. Such a scenario implies
that the hard and soft X-ray fluxes emitted by solar flares are
generally related, as was first described by \citet{Neup68}. Since
HXR and microwave emissions are produced by non-thermal electrons
and SXR are the thermal emission of a hot plasma, the Neupert effect
confirms that non-thermal electrons deliver energy spent on plasma
heating. Thus the HXR emission is directly related to the flux of
the accelerated electrons whereas the SXR emission is related to the
energy deposited by the non-thermal electron flux.

This interpretation, however, leads to a number of problems and
questions concerning energy content and time relations between SXR
and HXR fluxes (see e.g. \citet{Den88},  \citet{Den93}, and
\citet{McTier99}).

There are several papers that investigate temporal dependencies
between the beginnings of SXR and HXR emissions. For example,
\citet{Mach86} and \citet{Sch89} reported frequent strong SXR
emission before the impulsive phase. Following these authors, on
average, the SXR emission precedes the onset of HXR emission by
about 2 min. \citet{Ver02} analyzed 503 solar flares observed
simultaneously in HXR, SXR and $H\alpha$. In more than 90\% of the
analyzed flares, an increase of SXR emission began prior to the
impulsive phase. On average the SXR emission starts 3 min before the
hard X-ray emission.

An enhanced thermal emission preceding the onset of the hard X-rays
may be indicative of a thermal preheating phase prior to the
impulsive electron acceleration. Current-sheet models (e.g.,
\citet{Heyv77}) of solar flares predict a preheating phase.
\citet{Li87} discussed the preheating phase of solar flares
triggered by new emerging magnetic flux. They proposed four
different types of reconnections to explain the preheating as well
as impulsive phases of flares.

Another approach to this issue is multi-thread hydrodynamic modeling
of solar flares. \citet{Warr06} suggested that modeling a flare as a
bunch of independently heated threads may simulate precedence of the
SXR emission. The author successfully reproduced the temporal
evolution of high temperature flare plasma in the Masuda flare of
1992 January 13.

An alternative mechanism to electron beam-driven evaporation, namely
conducted driven evaporation, was developed lately by
\citet{Batt09}. They studied in detail the pre-flare phase of four
solar flares using imaging and spectroscopy from the \emph{RHESSI}
satellite. These authors explain the time evolution of the observed
emission for all analyzed events as an effect of saturated heat
flux.

The tendency of the soft X-ray flux to appear before hard X-rays
emission can be attributed also to the sensitivity threshold of the
hard X-ray detectors \citep{Den88}. At the beginning of the flare,
the energy flux may be below the detection threshold of HXR
emission.

In this paper we show, using unprecedented high sensitivity of the
RHESSI detectors \citep{Lin02} and a numerical model of flare, that
early SXR emission observed prior to impulsive phase could be fully
explained without any ad-hoc assumptions (at least for analyzed
event). All necessary energy to explain the soft emission could be
derived from observed HXR spectra. In Section 2 we describe the
analyzed event. Section 3 presents the details of the HXR spectra
fitting, numerical modeling and the results. The discussion and
conclusions are presented in Section 5.

\section{Observations}


For our work we selected RHESSI observations obtained without the
activation of the attenuators in order to prevent discontinuities in
fitting parameters.  A solar flare with a simple single-loop
structure was then chosen for convenience in numerical modeling. The
investigated flare occurred in the southwest hemisphere in active
region NOAA 10126 (S23E69) on 2002 September 20. It was classified
as an M1.8 \textit{GOES} class flare. \textit{GOES} X-ray
light-curves of the flare are shown in Figure 1 (top panel).
\textit{GOES} (1-8 ~\AA) flux has a background level of $1.04
\times10^{-6} ~~W/m^{2}~$ ~(C1.04). The SXR emission started to
increase slowly at 09:18:15 UT and showed two local maxima at
09:21:00 UT and 09:22:30 UT. It reached its maximum at 09:28:30 UT.
Harder (0.5-4 ~\AA) \textit{GOES} emission started to increase at
the same time as the softer one and peaked one minute earlier at
09:27:30 UT. It also showed two local, even more pronounced maxima
of the emission.

\begin{figure}[th!]
\includegraphics[angle=0,scale=0.4]{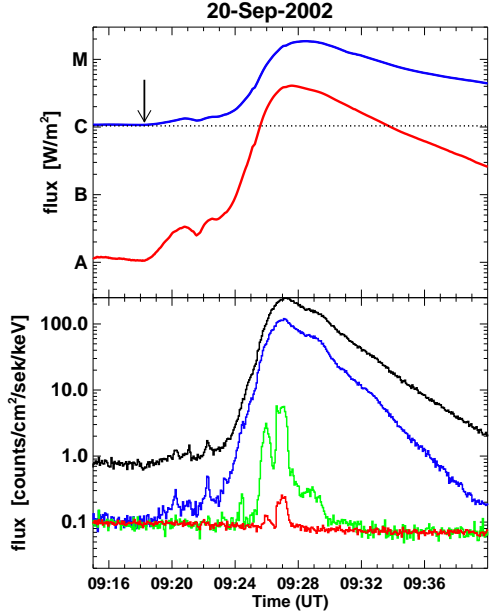}
\vspace{0.02 cm}\caption{Top panel: \textit{GOES} light curves in
two energy bands : 1-8 ~\AA ~(upper line)~and 0.5-4 ~\AA. The
horizontal dotted line represents the preflare background level
observed in soft channel. The arrow indicates the starting moment of
the numerical model. Bottom panel: \emph{RHESSI} light curves taken
in four energy bands (from top to bottom): 4-12, 12-25, 25-50 and
100-300 keV.}
\end{figure}

\emph{RHESSI} X-ray light curves of the flare taken in four energy
bands are shown in Figure 1 (bottom panel). The impulsive phase in
X-rays $ \ga 25\rm\, keV$ started at 09:25:24 UT and it had two
maxima around 09:26 UT and 09:27 UT, respectively. In 25-50 keV
energy range, a small spike of emission occurred between 09:24:16 UT
and 09:24:32 UT. It appeared also as a small hump on the both
\textit{GOES} light-curves. The X-ray emission below 25 keV started
to rise simultaneously with \textit{GOES} emission. Three local
peaks of emission are presented in these energy bands at the moments
of SXR local maxima.

Images obtained using \emph{RHESSI} data with the CLEAN imaging
algorithm revealed SXR emission at 6-12 keV and intermediate energy
emission 12-25 keV coincident with the flare location. These
observations indicate that SXR emission recorded by GOES in the
early phase of flare came from analyzed event.
\begin{figure}[th!]
\includegraphics[angle=0,scale=0.7]{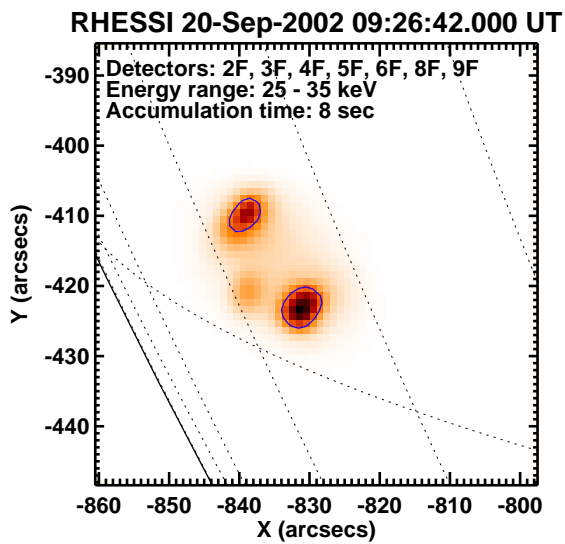}
\vspace{-0.5 cm}\caption{A 25-35 keV PIXON image with time
accumulation from 09:26:42 UT to 09:26:50 UT, at maximum impulsive
phase. An isocontour corresponds to 30\%  of the maximum flux. }
\end{figure}
Images obtained with the PIXON imaging algorithm \citep{Hur02}
showed a single flare loop (see Figure 2). This image allows us to
define geometrical proportions of the loop. The cross-sections of
the loop $S=1.13\times 10^{17} cm^2$ was estimated as an area within
a level equal to 30\% of the maximum flux in the 25-35 keV energy
range. Half-length of the loop $L_0=9.5\times 10^8 cm^2$ was
estimated from the distances between the centers of gravity of the
footpoints, assuming a semi-circular loop shape.

Unfortunately \emph{SOHO}/EIT telescope observed this AR before an
after the flare. However images obtained with 195 \AA ~ filter after
event at 09:47:59 and 9:59:59 UT confirmed the single loop structure
of the flare.

\section{Calculations - modeling of the heating of the loop}

The \emph{RHESSI} data were analyzed using RHESSI OSPEX package of
the SolarSoftWare (SSW) system. Data were summed over the front
segments of the seven detectors with detectors number 2 and 7
excluded. The  spectra were measured with 4 sec temporal resolution
in 158 energy bands from 4 to 300 keV.  We applied the energy widths
$dE = 0.3$ keV within the range 4-15 keV, $dE = 1.0$ keV in the
range 15-100 keV, and $dE = 5.0$ keV above 100 keV. The analyzed
spectra were corrected for pulse pile-up, decimation and albedo
effects. We used full 2D detector response matrix to convert input
photon fluxes to count rates. Before fitting spectra we removed the
averaged non-flare background spectra. For energies below 50 keV,
the background spectra were accumulated and averaged from pre-flare
period between 09:00 and 09:06 UT. For energies above 50 keV we used
a linear interpolations between the time intervals before and after
the impulsive phase.

In our model we used spectra taken after 09:18:15 UT when the SXR
emission started to increase. We used 4 seconds-long time bins, but
after background subtraction we increased the accumulation time in
the period from 09:18 UT to 09:23 UT to keep the positive counts
rates in most of the energy bins in the 4-20 keV range.

The spectra were fitted with single temperature thermal plus
thick-target models (vth + thick). The thermal model was defined by
single temperature and emission measure of the optically thin
thermal plasma, and is based on the X-ray continuum and line
emission calculated by the CHIANTI atomic code (\citet{Dere97},
\citet{Landi06}). The element abundances are based on the coronal
abundances of \citet{Feldman00}. The thick target model was defined
by the total integrated electron flux $N_{nth}$, the power-law index
of the electron energy distribution $\delta$, and the low energy
cut-off of the electron distribution $E_c$. Figure 3 shows the fit
of the spectrum, accumulated between 09:20:04 UT and 09:20:12 UT. In
addition to the thermal component, there is an important power law
shape for emission between 15 keV and 20 keV which can be recognized
and fitted as thick-target emission of the non-thermal electrons.

\begin{figure}[th!]
\includegraphics[angle=0,scale=0.5]{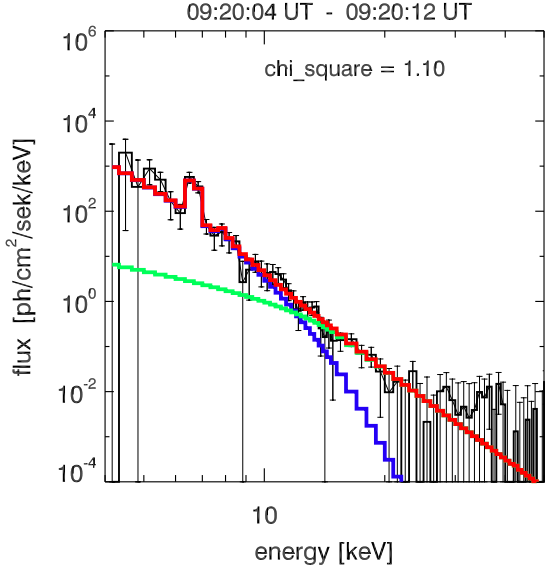}
\vspace{0.02 cm}\caption{\emph{RHESSI} spectral fit results for data
accumulated with a 8 sec time interval between 09:20:04-09:20:12 UT.
The spectrum was fitted with one temperature thermal model (blue
color) and thick target model (green color) with energy cut-off
$E_c$ = 15.8 keV  (please see the text for more details). Total
fitted spectrum is shown in red color. The errors bars of the
observed spectra are also shown.}
\end{figure}

We assumed the electron beam-driven evaporation model of the solar
flare. Therefore, we used in hydrodynamic model of the analyzed
flare the non-thermal electrons beams derived from \emph{RHESSI}
spectra as the heating source via the Coulomb collisions. Heating
was modeled using the approximation given by \citet{Fis89}. In this
work we used
1D Naval Research Laboratory Solar Flux Tube Model code
by Mariska and his co-workers (\citet{Mar82}, \citet{Mar89}). This
code was slightly modified by us. We included: new radiative loss
and heating functions, the VAL-C model of the initial structure of
the lower part of the loop, and double precision calculations. All
the changes were tested by comparison of the results with the
results of the original model. For details see the paper by
\citet{Fal09}. An important problem which we meet during the
modeling of the flares using original NRL code was an insufficient
amount of the mater located in feet of the loops. To solve this
problem we applied the VAL C model of the solar plasma, extended
down using Solar Standard Model data. It was done solely in order to
obtain big enough storage of the matter. All other aspects of the
NRL model of the chromosphere are unchanged (radiation is
suppressed, optically thick  emission is not accounted for, and no
account is taken of neutrals, the net conductivity flux is
negligible).

We modeled the evolution of the analyzed flare as follows: an
initial, quasi-stationary pre-flare models of the flaring loop was
built using geometrical and thermodynamic parameters estimated from
\emph{RHESSI} and \emph{GOES} data. These initial parameters of the
flaring loop were as follows: semi-length $9500 ~km$, radius $1900
~km$, pressure in feet $22 ~dyn/cm^2$. Small volumetric heating was
used to keep this model in the quasi stationary state on the
pre-flare level of activity before the start of the non-thermal
heating. Then we started to model the heating of the loop by
non-thermal electrons adding a dose of energy and calculating the
resulting \emph{GOES} flux. We used thick target parameters
$N_{nth}$,~ $\delta$ ~ and ~ $E_c$~ obtained from fits of
consecutive \emph{RHESSI} spectra for each time-step as input into
the Fisher's heating function. However, an acceptable fit can be
obtained for all ~ $E_c$~ values in the range ~5-30 keV for all
following spectra. Because the low energy cut-off determines an
amount of energy delivered to the loop, this non-uniqueness could be
limited using an independent energetic condition, like observed 1-8
\AA ~GOES flux. Indeed, for each time step we adjusted ~ $E_c$~
values in order to achieve conformity of the observed and modeled
fluxes in \emph{GOES} 1-8 \AA ~band. Such use of GOES 1-8 \AA ~flux
put important limitation on the allowed low energy cut-off values
and reduces importantly the non-uniqueness problem.

Fig. 3 shows an example of the \emph{RHESSI} fitted spectrum where
the low energy cut-off $E_c$ was adjusted to equalize synthesized
and observed \emph{GOES} fluxes in 1-8 \AA ~channel. The high energy
part of this spectrum was finally fitted with the following thick
target parameters: $\delta=7.37$,~ $N_{nth}=7.2\times 10^{33}~ \rm
electrons/sec$, and $E_c= 15.8 \pm 0.1 ~keV$. This non-thermal
electron beam contains total energy flux $F_{nth}=2.16 \pm 0.14
\times 10^{26}~\rm erg/sec$ and heated our loop during 8 seconds
between 09:20:04-09:20:12 UT. The heating increases the synthesized
\emph{GOES} class in this period from B1.93 to B2.11, which exactly
corresponds to the changes of the observed (with background removed)
\emph{GOES} flux. Given errors of $E_c$ and $F_{nth}$ were derived
by fitting the \emph{GOES} 1-8 \AA ~flux with an accuracy of 0.01
W/m2. Thermal component of the spectrum was fitted with emission
measure $EM= 4.25 \times 10^{46} cm^{-3}$ and temperature $T_e= 14.4
MK$. Obtained temperature is consistent with temperature obtained
from GOES data ($11.0 MK$). We fitted also the spectra with the
purely thermal model, but obtained temperature of $20.3 MK$ seems to
be quite high with respect to the GOES one and makes us more
confident in the interpretation with the presence of the non-thermal
electrons well before the impulsive phase.

\begin{figure}[th!]
\includegraphics[angle=0,scale=0.5]{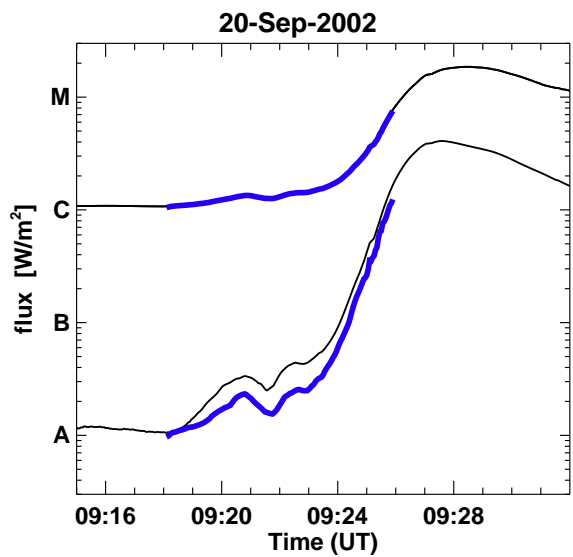}
\vspace{-0.5 cm}\caption{Comparison of the observed (black thin
lines) and calculated (thick blue line) \emph{GOES} fluxes in 1-8
\AA ~(upper curves) and 0.5-4. \AA ~(lower curves) energy bands
during initial phase of the analyzed flare.}
\end{figure}

The final result of our modeling is presented in Figure 4. The
synthesized  \emph{GOES} 1-8 \AA ~ light-curve closely follows the
observed one. The correspondence between observed and calculated
fluxes in the case of the 0.5-4 \AA ~band is not so ideal. The
inconsistency could be attributed to crude estimation of the initial
loop conditions, errors in \emph{RHESSI} spectra restoration, and
the simplicity of our model. However, because  the variations of the
calculated 0.5-4 \AA ~light-curve did not differ too much from the
observed variations, it means that our model simulate the main
physical processes in right way. As a result we fully restored the
observed slow increase of the SXR flux recorded before impulsive
heating using only non-thermal electrons characteristics derived
from the observed HXR spectra. Thus we confirmed that in the
analyzed flare variations of the SXR and HXR fluxes are consistent
with the Neupert effect.


\begin{figure}[th!]
\includegraphics[angle=0,scale=0.4]{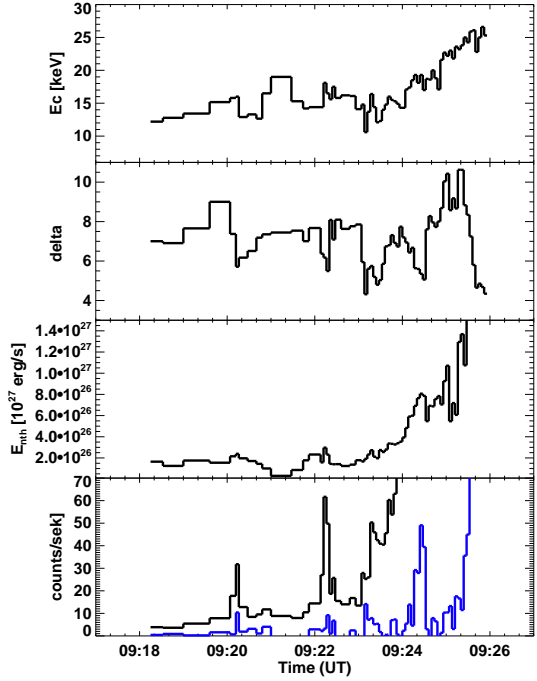}
\vspace{0.02 cm}\caption{Time evolution of the thick target
parameters and \emph{RHESSI} fluxes before the start of impulsive
phase. From top to bottom: $N_{nth}$,~ $\delta$, ~ $E_c$ ~ and
observed fluxes at 12-25 and 25-50 keV.}
\end{figure}

Figure 5 presents time variations of the electron beam (\emph{thick
target}) parameters and \emph{RHESSI} fluxes before the onset of the
main bulk of hard X-ray emission at 09:25:24 UT. Non-thermal
electron distributions, characterized by these parameters, delivered
to the loop an amount of energy sufficient to produce the slow
increase of the \emph{GOES} emission observed before the impulsive
phase. Low energy cut-off $E_c$ changed during this period between
12 keV and 24 keV. Electron spectral index $\delta$ varied between
4.5 and $\simeq$ 10, while total electron energy flux ranges from $
3.3 \times~10^{25}~\rm ergs/sec$~to $ 1.4 \times~10^{27}~\rm
ergs/sec$. Obtained variations of the $E_c$ reflect probably
temporal variations of the processes in primary energy source
region. On the other side obtained variations of the index $\delta$
show a clear general pattern of the soft-hard-soft spectral
evolution (see e.g. \citet{Gri04}). Additionally, all peaks in the
12-50 keV energy range are related to local increases in energy flux
of the non-thermal electrons and so to local increases in heating of
the loop. The two most pronounced peaks (and heating increases) at
09:20:16 UT and 09:22:16 UT were manifested as local flux
enhancements on both \emph{GOES} light-curves. The maximum
evaporation speed, achieved in a early phase of the simulation, was
equal to $130 ~km/s$. The maximum temperature, density, and pressure
obtained at the loop apex during the modeled period were equal to
$13.5 ~MK$, $1.3 \times 10^{11} ~cm^{-3}$, and $462~ dyn/cm^{2}$,
respectively.

\section{Conclusions}

The start and quick increase of HXR $ \ga 25\rm\, keV$ flux (defined
as an impulsive phase of flare) is interpreted as an indication of
the injection of the non-thermal electrons into the flaring loop and
a beginning of the plasma heating by these electrons. However, often
the SXR emission starts a few minutes earlier than the HXR which
raises a question about the heating source in very early stage of
the flare. Numerous solutions were proposed of which heating by
thermal conduction and Multi-Stranded Loop model are the most well
known. We showed that it is possible to fit soft (\textit{GOES}) and
hard (\emph{RHESSI}) X-ray emissions of a solar flares well before
beginning of the impulsive phase without any additional heating
besides the heating by non-thermal electrons. This was made possible
because of the unprecedented high sensitivity of the \emph{RHESSI}
detectors which are able to measure very low hard X-ray flux early
in the flare. Part of the emission, mainly in the energy range $< 25
keV$, is non-thermal in nature and indicates the presence of
non-thermal electrons. In this case of a M1.8 GOES class flare, the
non-thermal electron energy fluxes of the order of
$10^{26}~ergs/sec$, derived under the thick target interpretation,
fully explains the required heating of the plasma and resulting
increase in SXR emission.

The main limitations of our work are: crude estimation of the
initial physical and geometrical parameters of the loop, errors in
restoration of the RHESSI spectra and in GOES calibration, relative
simplicity of the 1D hydro-dynamical model and single loop
approximation of the event. While the synthesized GOES 1-8 \AA
~light-curve closely follows the observed one, all limitations
mentioned above caused small differences between observed and
calculated fluxes in the 0.5-4 \AA ~band.

Various authors considered classical heat conduction from the
loop-top energy reservoir. \citet{Batt09} fitted (\emph{RHESSI})
spectra of four flares with a purely thermal emission and claim that
at least for some flares electron beam heating does not work. In a
case of flare analyzed by us purely thermal spectra seems to give
too high temperatures ($> 20 ~MK)$ comparing to those obtained from
\emph{GOES} data. Additionally, even if the hard X-ray spectra
observed with \emph{RHESSI} provide no evidence for non-thermal
particles this does not necessarily means that the electrons don't
exists. As it was pointed e.g. by \citet{BH09} the non-thermal hard
X-ray emission associated with electron beam could be below
\emph{RHESSI's} level of detection. These authors estimated that in
the case of analyzed event the non-thermal X-ray emission produced
by an electron beam of sufficient energy flux to heat chromospheric
plasma to the temperature and emission measure observed by
\emph{RHESSI} is below the observed background level. It is possible
that for some flares heat conduction and for others beam driven
evaporation works on the early phase of flare evolution. In our case
the heating of the matter caused by non-thermal electrons is 2-4
orders higher then local conductive flux. \citet{Sto08} presented
analytic predictions of the X-ray $EM$ and $T_e$ expected in single
and filamented flare loops for both mechanisms of evaporation and
have tested these against real data consisting of 18 \emph{RHESSI}
microflares. Their results suggest beam heating in filamented loops
to be in agreement with data. The peak temperatures were consistent
with both single loop and multi-thread heating. The observed
emission measures were mostly compatible with beam driving for a
number of threads, but for 2 events EM were also compatible with
single loop model.

Our results extend the standard model of SXR and HXR relationship to
the early phases of solar flares and thus expands the number of
flares consistent with the Neupert effect. These results also
indicates that the process of electrons acceleration appears during
the early stage of the flare, well before the impulsive phase. This
was confirmed e.g. by \citet{Asai06} who reported detection of the
coronal non-thermal emission during the pre-impulsive phase of the
X4.8 flare on 2002 July 23.

Our method of adjustment of the low energy cut-off $E_c$ in order to
equalize synthesized and observed \emph{GOES} fluxes in 1-8 \AA
~channel can be considered as a new method of $E_c$ determination.

\section{Acknowledgments}
The authors acknowledge the \emph{RHESSI} consortium. This work was
supported by the Polish Ministry of Science and Higher Education,
grant number N203 022 31/2991 and by the European Community's
Seventh Framework Programme (FP7/2007-2013) under grant agreement
n\degr~ 218816 (SOTERIA).

\clearpage

\end{document}